# A linear phase evolution model for reduction of temporal unwrapping and field estimation errors in multi-echo GRE

Joseph Suresh Paul, Sreekanth Madhusoodhanan

Medical Image Computing and Signal Processing Laboratory, Indian Institute of Information Technology and Management- Kerala, Trivandrum – 695581, India

j.paul@iiitmk.ac.in

**Abstract**

This article aims at developing a model based optimization for reduction of temporal unwrapping and field estimation errors in multi-echo acquisition of Gradient Echo (GRE) sequence. Using the assumption that the phase is linear along the temporal dimension, the field estimation is performed by application of unity rank approximation (URA) to the Hankel matrix formed using the complex exponential of the channel-combined phase at each echo time. For the purpose of maintaining consistency with the observed complex data, the linear phase evolution (LPE) model is formulated as an optimization problem with a cost function that involves a fidelity term and a unity rank prior, implemented using alternating minimization. Itoh's algorithm applied to the multi-echo phase estimated from this LPE model is able to reduce the unwrapping errors as compared to the unwrapping when directly applied to the measured phase. Secondly, the improved accuracy of the frequency fit in comparison to estimation using weighted least-square regression (wLSR) and penalized maximum likelihood (pML) is demonstrated using numerical simulation of field perturbation due to magnetic susceptibility effect. It is shown that the field can be estimated with 80% reduction in mean absolute error (MAE) in comparison to wLSR and 66% reduction with respect to pML. The improvement in performance becomes more pronounced with increasing strengths of field gradient magnitudes and echo spacing.

## 1. INTRODUCTION

The relationship between the phase in gradient echo (GRE) acquisition and the local field variations caused by the susceptibility sources (1, 2) has enabled the clinical utility of deriving information on tissue iron content and quantitative estimation of susceptibility maps.The phase evolution inaGRE sequence is dependent onbothtotal field perturbation and echo time (TE). However, as the phase wraps and signal-to-noise ratio (SNR) vary with TE, a robust estimation of the total field perturbation is essential to improve the accuracy ofother quantitative maps derived from the original phase measurements such as the Quantitative susceptibility mapping (QSM) (3-8), perfusion and venous oxygen saturation (9-11). As the susceptibility under investigation is often on the order of ppm, error in the field map can complicate a meaningful measurement of susceptibility. Moreover, the ill-posed nature of QSM and sensitivity to error propagation further imposes stringent requirements on the accuracy of the input field map.

A number of methods are available in the literature to estimate the field from the measured multi-echo phase. A common approach is to compute the phase difference between two successive echoes and divide by the echo time difference (12-14). Since this approach do not account for the noise variation, the estimated field maps are very noisy. To address the noise effects, field map estimation using linear regression followed by median and Gaussian filtering was introduced (15). However this introduces undesired smoothing into the field estimate. Alternatively, a non-uniform echo spacing acquisition for field estimation was proposed by Aksit et al(16). Here the first two scans are performed with small echo spacing to avoid phase wrapping and the third with a higher echo time to improve the accuracy of the linear fit while maintaining the same acquisition time as that of the dual-echo acquisition. Succeeding to this, linear fitting of the temporal unwrapped phase was used to extract the frequency fit (17). However, this depends on the success of the temporal unwrapping. Later more sophisticated methods using maximum likelihood (ML) technique was proposed to overcome these limitations(18). Here regularized least square cost function is used without the explicit constraint of linearity. Since the cost function involves ML estimation, no temporal phase unwrapping is required. The noise characteristics in the estimated field are controlled using a penalized-likelihood cost function that involves an additional smoothness by adjusting the regularization parameter. Here the selection of the regularization parameter plays a key role in controlling the smoothness in the field estimate. Although penalized ML approach do not require prior phase unwrapping, a basic limitation of this approach is that it does not treat the true absolute phase as an unknown; while estimating the field directly using a non-convex optimisation that uses the original wrapped phase in its implementation. Alternatively, the proposed method works by iteratively modifying the phaseto satisfy the temporal linear phase evolution (TLPE)constraint; and thereby minimize the errors introduced by temporal unwrapping. This is based on the fact that with linear phase evolution, the frequency shift is independent of echo time. Consequently from Kronecker's theorem (19, 20), the Hankel matrix constructed from complex exponential of the phase shifts at each uniformly sampled (equi-spaced) echo will ideally possess unity rank. To maintain consistency with the complex data, a fidelity term is added along with the TLPE prior. This is formulated as an optimization problem and implemented using alternating minimization. The principal values of each echo phase shift obtained as the solution to this problem, are then unwrapped using Itoh's algorithm. The rank approximated principal phase values when unwrapped, results in lower unwrapping errors.

The proposed method gives accurate phase estimates at each echo and reliable field estimates in comparison to weighted least-squares regression (wLSR) and penalized ML (pML. For field map estimation, the relative reduction in RMSE for LPE+wLSR is found to be 98 % with respect to wLSR and 95% with respect to PML.

## 2. MATERIAL AND METHODS

## 2.1 Multi-echo GRE signal model

The channel-combined signal obtained using GRE sequence with an echo time difference $\Delta TE$ can be modelled as

$$g_j^l = \delta_j^l exp(-i(\omega_j(TE_0 + l\Delta TE) + v_j)) + \epsilon_j, \quad [1]$$

where $\delta_j^l \triangleq \rho_j exp(-(TE_0 + l\Delta TE)/T_{2j}^*)$ is the temporal signal intensity of the $T_2^*$-weighted GRE sequence at the $j'$th location, $\epsilon_j^l$ denotes the complex zero mean noise component with variance $\sigma^2$, $\omega_j$ is the field perturbation(offset frequency) due to the susceptibility effect and $v_j$ is the receiver phase offset. The signal intensity is assumed to decay exponentially with a magnitude determined by the proton density $\rho_j$, and rate dependent on thetime constant $T_{2j}^*$. The echoes are assumed to be equally spaced at an interval $\Delta TE$, $l$ is the echo index, $l = 0, \dots, L - 1$, $TE_0$ is the echo time at $l = 0$.

## 2.2 Magnitude Estimation

To simplify the problem, we first determine the ML estimate for the magnitude vector $\boldsymbol{\delta}^l = [\delta_1^l, \delta_2^l, \dots, \delta_N^l]$ to determine the true phase based on the LPE constraint. Under the assumption of known valuesof $\boldsymbol{\omega} = [\omega_1, \omega_2, \dots, \omega_N]$ and $\boldsymbol{v} = [v_1, v_2, \dots, v_N]$, the joint likelihood function is

$$\log p(\mathbf{g}; \boldsymbol{\delta}) = \sum_{l=1}^{L} \log p(\mathbf{g}^l; \boldsymbol{\delta}^l),$$

[2]

$$\equiv \frac{1}{2\sigma^2} \sum_{j=1}^{N} \sum_{l=1}^{L} \left| g_j^l - \delta_j^l \exp\left(-i(\omega_j(TE_0 + l\Delta TE) + v_j)\right) \right|^2,$$

where '≡ ' denotes equality to within constants independent of **δ**. Minimizing the negative log-likelihood for each channel with respect to $\boldsymbol{\delta}^l$ gives

$$\boldsymbol{\delta}^l = Real(\mathbf{g}^l)\cos(\widehat{\boldsymbol{\psi}}^l) - Imag(\mathbf{g}^l)\sin(\widehat{\boldsymbol{\psi}}^l) \qquad [3]$$

To a first approximation, the phase estimates $\widehat{\boldsymbol{\psi}}^l$ to be used in [3] can be taken as $\widehat{\boldsymbol{\psi}}^l = arg(\mathbf{g}^l)$; which is the original acquired phase. The updated magnitude is then used to re-estimate the phase using the proposed phase optimization procedure (See Phase recovery using LPE constraint). By alternatively updating the magnitude and phase in this manner, the accuracy of field estimates can be improved especially in regions with low SNR.

## 2.3 Phase Unwrapping

Given that the echoes are equi-spaced at ΔTE, the sampling theory implies that the maximum absolute phase gradient that can be measured and unwrapped accurately is $\pi$ radians (21). It is useful to know the maximum acquisition time and the minimum number of echoes possible such that no aliasing occur. With the known magnitude samples, the phase can be recovered by application of Nyquist sampling theorem to the uniformly sampled complex exponential of the unknown phase measured at successive echo times. Thus for the signal $h \triangleq exp(-i(\omega t + v)))$ with the true phase $\varphi(t) \triangleq \omega t + v$ and maximal frequency $\left(\frac{\omega}{2\pi}\right)$ cps, the signal recovery is possible from the measurements at a series of samples spaced at $\frac{\pi}{\omega}$ sec apart. This implies that with linear phase evolution, the phase recovery is possible if

$$\omega \Delta TE < \pi. \qquad [4]$$

Assuming that this is satisfied, the next step in the process of phase recovery is unwrapping. With the true phase represented by $\varphi(t)$, the measure phase $\psi(t)$ is expressed in terms of the wrapping operator W. In the equation shown below, the wrapped phase is determined by

$$\psi(t) = \varphi(t) + 2\pi k(t) = W\{\varphi(t)\}, \qquad [5]$$

where $k(t)$ is an integer function that forces the true phase, $\varphi$ to be in the range $-\pi < \psi < \pi$ radians. The unwrapping process can be thought of as the inverse of the wrapping operation. Choices must be made for the appropriate values of $k$, such that the phase estimate is as accurate as possible. The result that Itoh found was that the true phase $\varphi$ can be recovered by integrating the wrapped differences of the wrapped phase samples. In the one-dimensional case, Itoh's algorithm describes the solution of this problem by integration of the phase gradients from sample to sample. The true unwrapped phase is calculated from

$$\varphi(m) = \varphi(0) + \sum_{l=0}^{m} W[\Delta(W[\varphi(l)])], \qquad [6]$$

where $\varphi(0)$ is the initial phase of a reference point and $\Delta(l)$ is the wrapped phase difference defined as $\Delta(l) \triangleq W[\psi(l+1) - \psi(l)]$. The wrapped phase differences are computed by differentiation of the result from the temporal phase measurements, and wrapping this difference back to the original $\pi$ interval. This solution implies that the original signal is sampled at a sample rate sufficiently high, so that the "true" phase differences compared to phase jumps introduced by the wrapping process between neighboring points is always in the range of $-\pi/2 \leq \varphi(l) - \varphi(l-1) < \pi/2$.

In the one-dimensional case, the solution $\varphi$, is the true phase if $\Delta[\varphi(l)]$ lies in the interval $-\pi < \psi < \pi$, and is otherwise an unwrapped phase estimate which is not exactly the same the true phase. This can lead to unwrapping errors. Therefore, when the phase is unwrapped, it is expected that the unwrapped phase will represent the true phase, if $\Delta(l) < \pi$. Only if this condition is true, the original phase can be accurately recovered.

### 2.4 Phase Recovery

From [1] the measured phase at $j'$th location can be expressed as

$$\hat{\psi}_j^l = W[\omega_j(TE^0 + l\Delta TE) + v_j + \eta_j^l], \qquad [7]$$

where $\hat{\psi}_j^l = arg(g_j^l)$ and $\psi_j^l = W[\omega_r(TE^0 + l\Delta TE) + v_j]$ represents the true wrapped phase. The variance of noise component $\eta_j^l$ is inversely dependent on the magnitude $\delta_j^l$ and directly on $var(\epsilon_j)$. Thus the complex exponential signal $h_j^l = h_j(TE^0 + l\Delta TE) \triangleq exp(-i(\omega_j(TE^0 + l\Delta TE) + v_j + \eta_j^l)))$ in the temporal dimension is uniformly sampled, but without linear phase due to the presence of $\eta_j^l$. Since $\hat{\psi}_j^l$ is the wrapped form of the phase of $h_j^l$, the periodic nature of the complex exponential signals also implies $h_j^l = exp(-i\hat{\psi}_j^l)$. Using the property that a Hankel matrix constructed from uniformly sampled complex exponential signal with time independent (constant) magnitude and linear phase will ideally have unity rank, a unity rank approximation of the Hankel matrix constructed from $h_j^l$, $l = 0, \ldots, L-1$; can be used to eliminate the phase noise and make the phase linear. In this form the true underlying complex exponential $f_j^l = exp(-i\psi_j^l)$. In vectorial form, a collection of the uniformly sampled complex exponential signals with non-linear phase is represented as

$$\mathbf{h}_j = \begin{bmatrix} h_j^0 \\ h_j^1 \\ \vdots \\ h_j^{(L-1)} \end{bmatrix}. \qquad [8]$$

Using the window length ($w$) that satisfies $2 \leq w \leq L$, $\mathbf{h}_j$ can be mapped to a 2D structured Hankel matrix $H_{h_j}$. Similar vectorized notations can be adopted to represent other measured and estimated quantities. Hereafter, we drop the subscript $'j'$ with the understanding that the remaining analysis applies separately to each voxel in the image. If the window length is chosen to be $w = (L+1)/2$, $H_h$ can be represented as

$$H_h = \begin{bmatrix} h^0 & h^1 & \cdots & h^{w-1} \\ h^1 & h^2 & \cdots & h^w \\ \vdots & \vdots & \ddots & \vdots \\ h^{L-w} & h^{L-w+1} & \cdots & h^{L-1} \end{bmatrix}. \qquad [9]$$

Since $H_h$ would be a square matrix with dimension $w \times w$, the singular value decomposition (SVD) can be applied for rank approximation with minimum number of flops (22, 23). With $H_f$ representing the rank reduced approximating matrix and $f_j$ the complex exponential signal vector recovered from $H_f$, the optimization problem can be viewed as a restricted low rank approximation (RLRA) (24, 25). Given the measured signal $\boldsymbol{g}$, the vectorial form of the magnitude signal at $j'$th location is

$$\boldsymbol{g}_j = \begin{bmatrix} |g^0| \\ |g^1| \\ \vdots \\ |g^{L-1}| \end{bmatrix}. \qquad [10]$$

A hankel matrix G can be constructed in the same way as the mapping from $h$ to $H_h$.

$$G = \begin{bmatrix} |g^0| & |g^1| & \cdots & |g^{w-1}| \\ |g^1| & |g^2| & \cdots & |g^w| \\ \vdots & \vdots & \ddots & \vdots \\ |g^{L-w}| & |g^{L-w+1}| & \cdots & |g^{L-1}| \end{bmatrix}. \qquad [11]$$

Likewise, D represents the hankel matrix mapped from magnitude signal $d$ at each location obtained through ML estimation.

Using the hankel matrices constructed separately for magnitude and phase, the restricted rank approximation problem for phase recovery is given by

$$\arg\min_{H_f} \sum_{j=1}^{N} \|D \circ H_f - G \circ H_h\|_F^2 ;$$

$$\text{subject to } \begin{cases} rank(H_f) = 1 \\ D(p,q) \circ H_f(p,q) - d(p+q) \circ f(p+q) = 0; \\ 0 \leq p, q \leq w - 1 \end{cases} \quad [12]$$

To solve this using alternating direction method of multiplier (ADMM) (26, 27), the cost function in [12] is now modified as

$$\arg\min_{H_f, f} \frac{1}{2} \|D \circ H_f - G \circ H_h\|_F^2 + \lambda \mathcal{R}(H_f);$$

$$\text{subject to } D(p,q) \circ H_f(p,q) - d(p+q) \circ f(p+q) = 0; \quad [13]$$

$$0 \leq p, q \leq w - 1,$$

where $\lambda$ is the regularization parameter and $\mathcal{R}(H_f)$ denote the indicator function:

$$\mathcal{R}(H_f) \triangleq \begin{cases} 0 & \text{if } rank(H_f) = 1 \\ \infty & \text{otherwise} \end{cases}, \quad [14]$$

The augmented Lagrangian form of [13] with scaled dual variable is

$$\mathcal{L}(H_f, f, u) = \|D \circ H_f - G \circ H_h\|_F^2 + \lambda \mathcal{R}(H_f) +$$

$$\frac{\rho}{2} \sum_{p,q=0}^{w-1} |D(p,q) \circ H_f(p,q) - d(p+q) \circ f(p+q) + u(p+q)|^2; \quad [15]$$

$$0 \leq p, q \leq w-1,$$

where $u$ is the Lagrange multiplier of the LPE constraint and $\rho$ is the penalty parameter. The ADMM steps for the intermediate stage $k$ to $k+1$, involve solving the subproblems:

$$H_f^{k+1} = \arg\min_{H_f} \mathcal{L}(H_f, f^k, u^k), \quad [16]$$

$$f^{k+1} = \arg\min_f \mathcal{L}(H_f^{k+1}, f, u^k), \quad [17]$$

$$u^{k+1}(p+q) = u^k(p+q) +$$
$$D(p,q) \circ H_f^{k+1}(p,q) - d(p+q) \circ f^{k+1}(p+q). \quad [18]$$

The ADMM implementation for solution of [16], [17] & [18] is included in the Appendix. To solve Eq. [12], we propose to use an efficient iterative temporal phase recovery algorithm as described in Algorithm 1.

**Algorithm 1**. Iterative Temporal Phase Recovery with LPE constraint
**Input:** Channel combined complex exponential signal
$g^l \triangleq g_j^l = [g]_{j,l}; l = 0,1, \ldots, L-1$.
**Initialization:** Iteration index $k = 1; w = \frac{L+1}{2};$

$$\begin{aligned}
&\text{Construct } H_h \text{ from } h^l \triangleq exp\left(-i\arg(g^l)\right); l = 0,1,\dots,L-1 \quad.\\
&\text{Construct G from } |g^l|; l = 0,1,\dots,L-1.\\
&\hat{\psi}^l = \arg(g^l); l = 0,1,\dots,L-1.
\end{aligned}$$

**Iteration:** do {
1. Obtain the ML updated magnitude $d^{l^{(k+1)}}$ from $\hat{\psi}^{l^{(k)}}$.
2. Construct $D^{(k+1)}$ from $d^{l^{(k+1)}}$.
3. Update for subproblem-1 :- Compute $\hat{H}_f^{k+1}$ using [A7].
   (a). $U\Sigma V = SVD(\hat{H}_f^{k+1})$.
   (b). convex projection to restore Hankel structure
   $H_f^{(k+1)} = Hankelize(U_1 \Sigma_1 V_1^H)$.
4. Obtain $\mathbf{f}^{(k+1)}$ using [A10].
5. Obtain $\mathbf{u}^{(k+1)}$ using [18].
6. $\hat{\boldsymbol{\psi}}^{(k+1)} \triangleq arg(\mathbf{f}^{(k+1)})$.
$k = k+1$
} till convergence or maximum number of iteration.

## 2.5 Temporal phase unwrapping under sub Nyquist condition

After phase recovery with rank approximation, the principal angles of the linear phase are generated. Since $\boldsymbol{\psi}_j$ is wrapped, temporal phase unwrapping need to be performed prior to field estimation. In general the temporal unwrapping fails due to two reasons :- 1) phase wrap exists in the initial echo, 2) Nyquist condition is not satisfied (i.e., $\omega\Delta TE > \pi$). The phase wrapping artifacts may exist in the first echo due to either a large receiver phase offset or higher $B_0$ field. Temporal unwrapping in case 1) would require additional steps that involve estimation of the receiver phase offset or spatial unwrapping of the phase image at $TE^0$.

In this section, we elaborate an algorithmic procedure to eliminate the temporal unwrapping error due to the sub-Nyquist condition when the underlying phase in linear. As discussed in the previous section, the optimization framework for rank approximation and reduction of higher order spectral energy using the LPE constraint guarantee a linear phase even in the presence of noise and receiver phase offset in the measured phase. However, this is not the case if the rank approximation were not performed apriori. Due to the linearity ensuing from rank approximation,the field estimation requiresonly true estimates of at least two sampling points along the temporal dimension. Of these, it is further assumed that the measured phase at $TE^0$ is wrap-free.In this situation, the unwrapped phase ($\varphi_j^1$) at the second echo can be estimated after introduction of an additional sample so that the Nyquist condition is satisfied. We denote $\psi_j^{0,1}$as the unknown phase at the pseudo sampling location (0,1) that corresponds to a fictitious acquisition time$(TE^0 + \frac{\Delta TE}{2})$.With the introduction of this new sample, $\psi_j^1$ can be unwrapped using the Itoh's algorithm in two steps.The first step consists ofunwrapping$\psi_j^{0,1}$. Application of Itoh's algorithm in this step leads to

$$\varphi_j^{0,1} = \begin{cases} \psi_j^{0,1} - 2\pi & if\ \psi_j^{0,1} - \psi_j^0 > \pi \\ \psi_j^{0,1} + 2\pi & if\ \psi_j^{0,1} - \psi_j^0 < -\pi \\ \psi_j^{0,1} & if\ |\psi_j^{0,1} - \psi_j^0| \leq \pi \end{cases}, \quad [19]$$

and

$$\hat{\varphi}_j^1 = \begin{cases} \psi_j^1 - 2\pi & if\ \psi_j^{0,1} - \psi_j^0 > \pi \\ \psi_j^1 + 2\pi & if\ \psi_j^{0,1} - \psi_j^0 < -\pi \\ \psi_j^1 & if\ |\psi_j^{0,1} - \psi_j^0| \leq \pi \end{cases}, \quad [20]$$

In the second step, Itoh's algorithm gives

$$\varphi_j^1 = \begin{cases} \hat{\varphi}_j^1 - 2\pi & if\ \psi_j^1 - \psi_j^{0,1} > \pi \\ \hat{\varphi}_j^1 + 2\pi & if\ \psi_j^1 - \psi_j^{0,1} < -\pi \\ \hat{\varphi}_j^1 & if\ |\psi_j^1 - \psi_j^{0,1}| \leq \pi \end{cases}, \quad [21]$$

Based on the above two steps, $\varphi_j^1$ can assume nine possible solutions as summarized in Table-1.

Table1: Two step unwrapping of $\psi_j^1$ with all possible solutions

| Step1 | | Step 2 | |
|---|---|---|---|
| $\Delta_1 = \psi_j^{0,1} - \psi_j^0$ | | $\Delta_2 = \psi_j^1 - \psi_j^{0,1}$ | |
| Condition | Unwrapped phase | Condition | Unwrapped phase |
| $\Delta_1 > \pi$ | $\varphi_j^{0,1} = \psi_j^{0,1} - 2\pi$ | $\Delta_2 > \pi$ | $\varphi_j^1 = \hat{\varphi}_j^1 - 2\pi = \psi_j^1 - 4\pi$ |
| | $\hat{\varphi}_j^1 = \psi_j^1 - 2\pi$ | $\Delta_2 < \pi$ | $\varphi_j^1 = \hat{\varphi}_j^1 + 2\pi = \psi_j^1$ |
| | | $|\Delta_2| \leq \pi$ | $\varphi_j^1 = \hat{\varphi}_j^1 = \psi_j^1 - 2\pi$ |
| $\Delta_1 < \pi$ | $\varphi_j^{0,1} = \psi_j^{0,1} + 2\pi$ | $\Delta_2 > \pi$ | $\varphi_j^1 = \hat{\varphi}_j^1 - 2\pi = \psi_j^1$ |
| | $\hat{\varphi}_j^1 = \psi_j^1 + 2\pi$ | $\Delta_2 < \pi$ | $\varphi_j^1 = \hat{\varphi}_j^1 + 2\pi = \psi_j^1 + 4\pi$ |
| | | $|\Delta_2| \leq \pi$ | $\varphi_j^1 = \hat{\varphi}_j^1 = \psi_j^1 + 2\pi$ |
| $|\Delta_1| \leq \pi$ | $\varphi_j^{0,1} = \psi_j^{0,1}$ | $\Delta_2 > \pi$ | $\varphi_j^1 = \hat{\varphi}_j^1 - 2\pi = \psi_j^1 - 2\pi$ |
| | $\hat{\varphi}_j^1 = \psi_j^1$ | $\Delta_2 < \pi$ | $\varphi_j^1 = \hat{\varphi}_j^1 + 2\pi = \psi_j^1 + 2\pi$ |
| | | $|\Delta\psi_j| \leq \pi$ | $\varphi_j^1 = \hat{\varphi}_j^1 = \psi_j^1$ |

From the last column of the Table-1, it is observed that out of the nine possible ways to unwrap $\psi_j^1$, there exist five uniquevalues for $\varphi_j^1$. This is due to the cumulative sum of phase rotations calculated in the two steps of unwrapping. These five cases can be uniquely identified by examining the ranges of $\Delta_1 \triangleq \psi_j^{0,1} - \psi_j^0$ and $\Delta_2 \triangleq \psi_j^1 - \psi_j^{0,1}$. For example in the two step unwrapping, if $\Delta_1 > \pi$, $2\pi$ is subtracted from the wrapped phase and in the next step if $\Delta_2 < -\pi$, $2\pi$ is added to the phase obtained in the first step. This result in an unwrapped phase $\varphi_j^1 = \psi_j^1$. The same unwrapped value will be obtained if either $\Delta_1, \Delta_2 \in [-\pi, \pi]$ or $\Delta_1 < -\pi$ and $\Delta_2 > \pi$. Other cases correspond to $\Delta_1 > \pi, \Delta_2 \in [-\pi, \pi]$ or $\Delta_1 \in [-\pi, \pi], \Delta_2 > \pi$ resulting in $\varphi_j^1 = \psi_j^1 - 2\pi$ and $\Delta_1 < -\pi, \Delta_2 \in [-\pi, \pi]$ or $\Delta_1 \in [-\pi, \pi], \Delta_2 < -\pi$ resulting in $\varphi_j^1 = \psi_j^1 + 2\pi$.

Among the five possible solutions for $\varphi_j^1$, the choices $\psi_j^1 + 4\pi$ and $\psi_j^1 - 4\pi$ apply only for large values of $\omega$. However, such high field do not exist in practical conditions and too high $\Delta TE$ values are also not used. For example, in a human brain that is dominated by water with tissue susceptibility $\chi = -9.05$ ppm, and grey white matter susceptibilities in the range $-9.2$ to $-8.8$ ppm, the forward field estimation yields values in the range $-300 \leq \omega \leq 300$ rad/s at 3.0 T and $\Delta TE$ limited in the range 1-12 ms. Hence the choices with $\pm 4\pi$ addition can be conveniently excluded. However, in cases where such possibilities arise, the number of intermediate sampling points should be more than one and the number of possible solutions would then increase. After arriving at the possible solutions, we apply the Itoh's algorithm as outlined in the flowchart shown in Fig. 1.

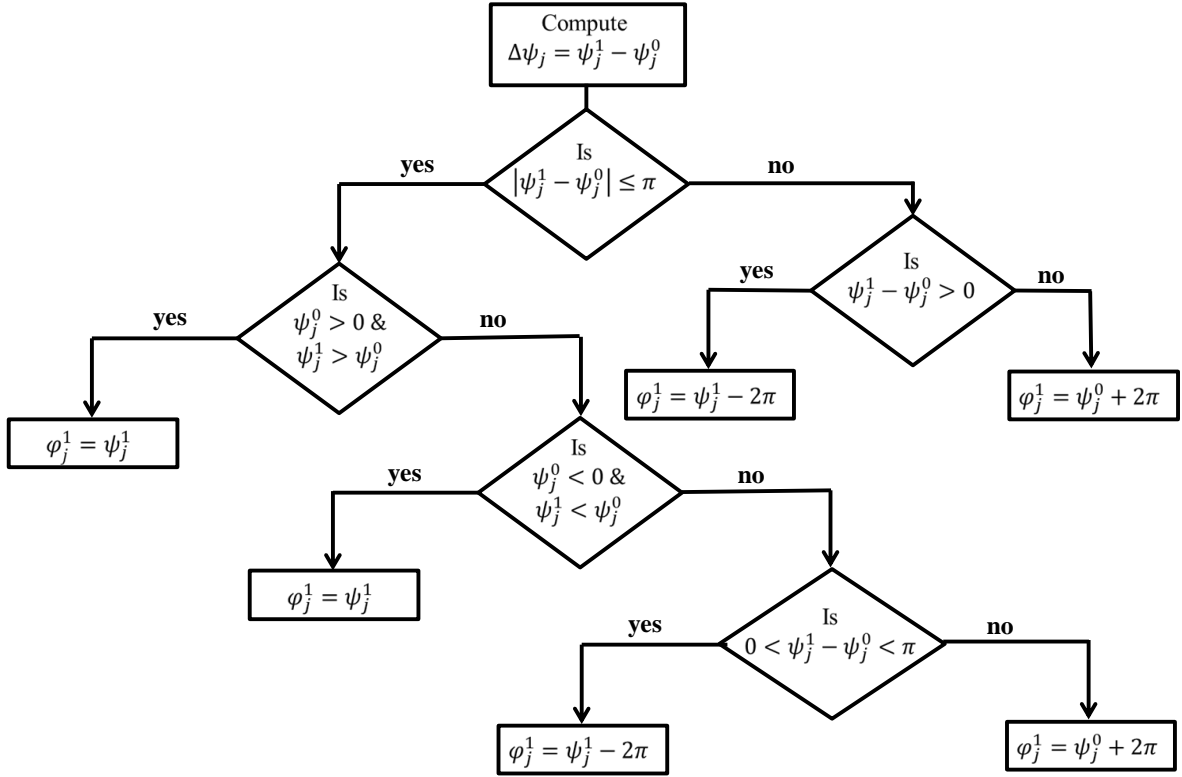

**Figure 1:** Flowchart for temporal phase unwrapping under sub-Nyquist condition.

## 3. RESULTS

### 3.1. One dimensional numerical experiments

#### 3.1.1 Temporal phase unwrapping errors

To simulate the wrapped phasealong temporal dimension, a one-dimensional uniformly sampled complex exponential signal having linear phase shift andexponentially decaying magnitude with decay rate determined by a prefixed value of $T_2^*$ is first generated. The parameters chosen are $L=7$, $TE^0 = 20\ ms$ and $\Delta TE = 12\ ms$. Once the ground truth signal with linear phase evolution is simulated, a complex noise component of known standard deviation is added to the complex exponential signal. A wrapped form of the temporal phase is then obtained from the principal angles of the noisy complex exponentials. Itoh's algorithm is then applied to unwrap the one-dimensional phase samples. Unwrapping error is measured by computing the absolute error between the unwrapped phase and the groundtruth. Fig .2 shows the plots of unwrapping errors in the $2^{nd}$ and $3^{rd}$ echo as a function of SNR and $\omega$. A key observation is that the unwrapping errors are much higher at low SNRs. The plots showing the dependence on $\omega$ are included in the bottom two rows of Fig.2.

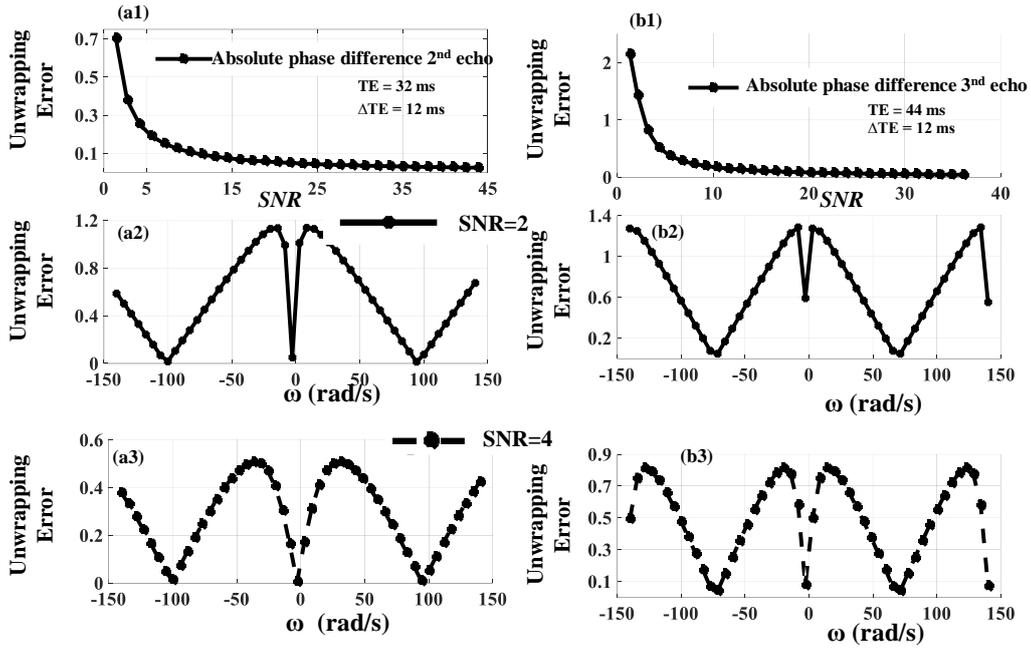

Figure2: Temporal phase unwrapping error expressed as absolute phase difference between the unwrapped phase (Itoh's algorithm) and the groundtruth linear phase. Left and right panels show unwrapping errors for $2^{nd}$ and $3^{rd}$ echo at 32 ms and 44 ms. (a1-b1) Plots of unwrapping error versus SNR . Bottom two rows show the unwrapping error as function of $\omega$ at two different SNRs. The maximum frequency is within the Nyquist limit; $\omega * \Delta TE \leq \pi$.

### 3.1.2 Unity rank approximation

Fig.3 illustrates the rationale for unity rank approximation of the Hankel matrix constructed from the uniformly sampled complex exponentials with linear phase shift. Top row shows (a) a sample Hankel matrix generated from uniformly sampled complex exponentials ($L = 7$), with linear phase shift and (b)sample singular value spectra for:- 1) without receiver phase offset and noise, 2) with receiver phase offset and without noise, 3) with receiver phase offset and noise. It is observed that for the first two cases where noise is absent,only the first singular value is non-zero and the rank is ideally unity. With complex noise addition, the rank of the Hankel matrix is observed to be full rank (all singular values are non-zero). The bottom panel (c) illustrates the empirical convergence of the rank approximation algorithmand accompanying reduction in the relative higher order spectral energy (rHSE)for different values of receiver phase offset and $\omega$.

### 3.1.3 Effect of rank approximation on the recovered phase

The effect of rank approximation on the reduction of unwrapping error is shown in Fig.4. The temporal phase unwrapping error is measured as the absolute phase difference between the groundtruth linear phase and the phase obtained by unwrapping the phase values updated using the rank approximation algorithm. It is observer that at low SNR, the unwrapping errors are reduced after application of rank approximation. The row-wise panels show the plots of unwrapping error versus SNR for three different $\omega$ values. As seen before, the unwrapping errors reduce with increase in SNR.

### 3.1.4 Effect of rank approximation on field estimation

Field estimation is performed from the simulated uniformly sampled complex exponentials using wLSR, pML and LPE+wLSR. wLSR method uses the temporalunwrapped phase for field estimation whereas pML method uses the wrapped principal angles. In the LPE+wLSR method, the field is estimated using wLSR applied to the temporal unwrapped phase obtained after rank

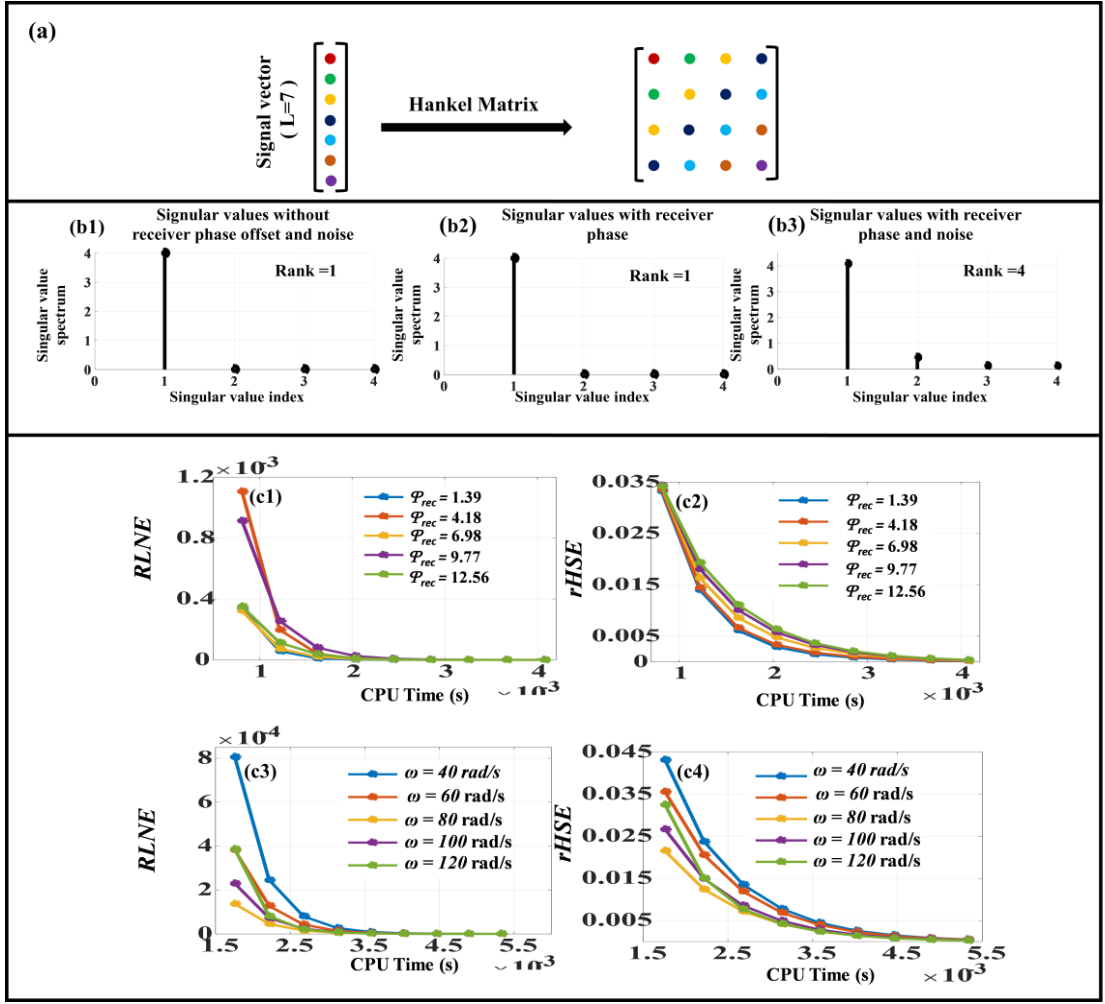

**Figure3:** Need for unit rank approximation of Hankel matrix with cross diagonals comprising of successive samples of a uniformly sampled complex exponential with linear phase and constant magnitude (magnitude is taken as unity). Top panel (a) Hankel matrix constructed from a signal vector with L=7 samples. (b1-b3) Singular value spectrum for the Hankel matrix constructed with (b1) uniformly sampled signal with no receiver phase offset and noise, (b2) uniformly sampledsignal with receiver phase offset and no noise component, (b3) uniformly sampledsignal with receiver phase offset and noise. In the first two cases, only the first singular value is non zero and rank of the Hankel matrix is ideally unity. With complex noise addition, the rank of the Hankel matrix is observed to be 4 (Full rank). The bottom panel (c) illustrates the convergence of the optimization procedure for unit rank approximation and reduction in relative higher order spectral energy (rHSE). (c1) plots of RLNE versus iterations (CPU time) for different receiver phase offsets, (c2) plots of rHSE versus CPU time for different receiver phase offsets, (c3) plots of RLNE versus CPU time shown for different $\omega$, (c4) plots of rHSE versus CPU time shown for different $\omega$.

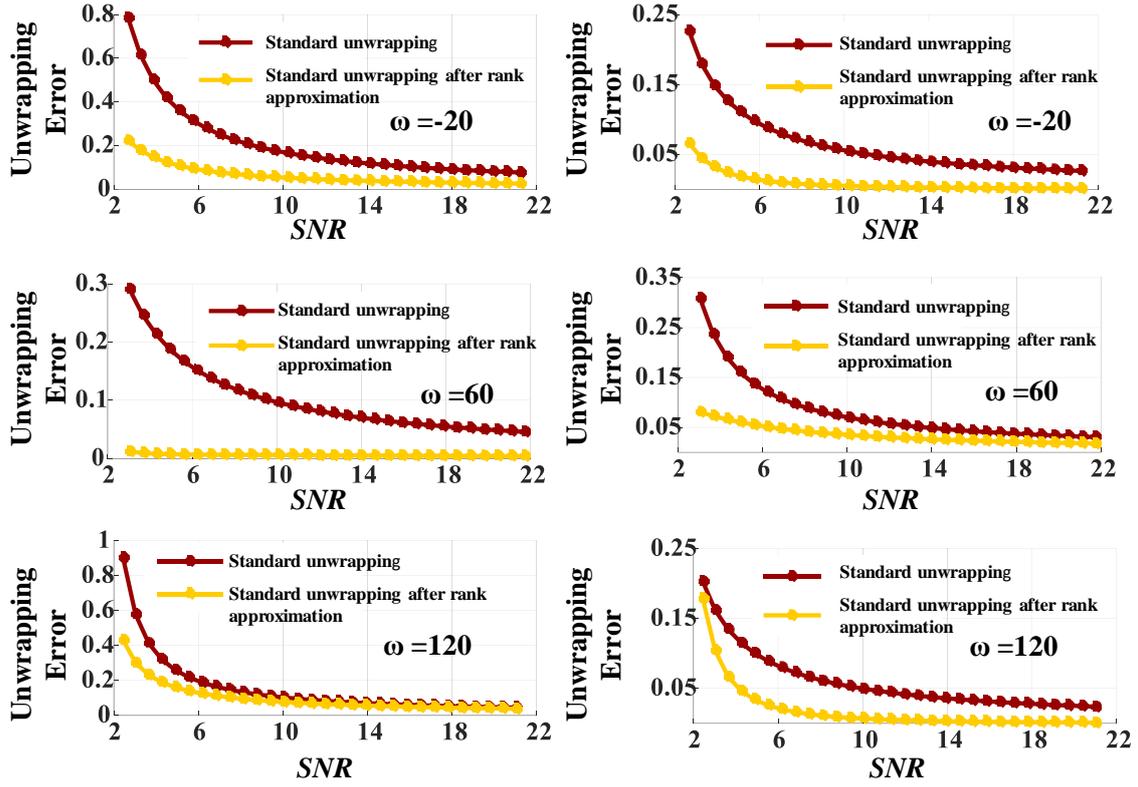

Figure 4: Temporal phase unwrapping error expressed as absolute phase difference between the unwrapped phase and the groundtruth linear phase. Left and right panels show unwrapping errors for 2nd and 3rd echo at 32 ms and 44 ms. Row wise panelsshow plots of unwrapping error versus SNR for three different ω values. Each panel show the unwrapping error obtained for standard unwrapping (Itoh's algorithm) and standard unwrapping applied after rank approximation, using optimization with LPE constraint.

approximation with LPE constraint. To measure the field estimation error with receiver phase offset ($\upsilon$), the complex exponentials are simulated with different $\upsilon$ values in the range [$-2\pi$, $2\pi$]. Fig.5 shows the plots offield estimation error (the absolute difference between the estimated field and the groundtruth) versus $\upsilon$.It is observed that the field estimation errors are minimal for LPE+wLSRat all receiver phase offsets.

In another experiment, field estimation errors are shown as functions of SNR and $\Delta TE$ for a given values of $\upsilon$ and $\omega$ ($\upsilon = \pi$ rad and $\omega = 120$ rad/s).The top and bottom row in Fig.6 show the plots of field estimation

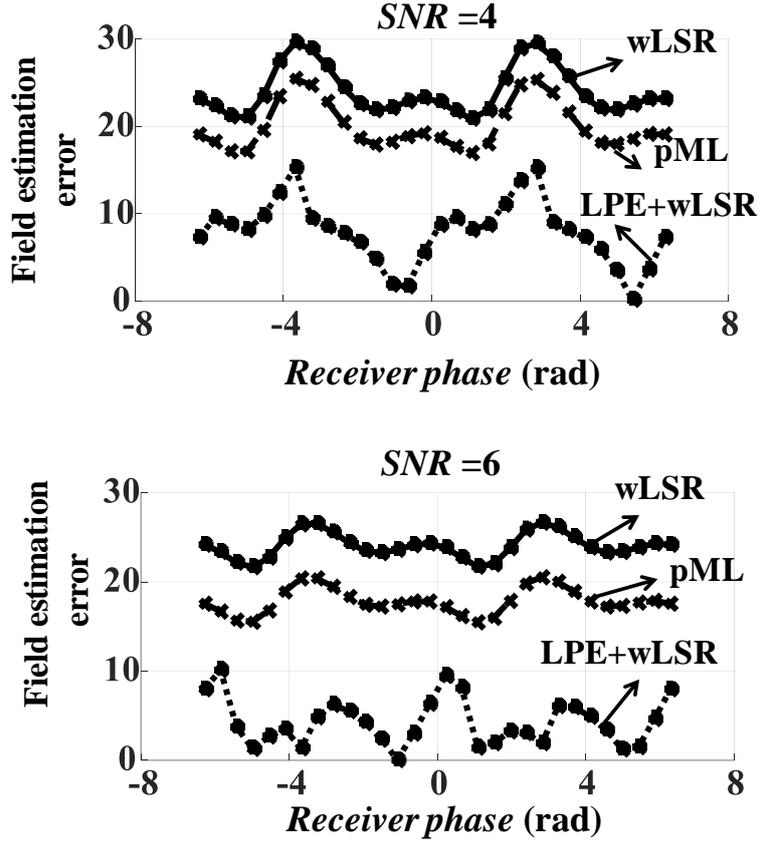

figure5: Field estimation error expressed as absolute difference between the estimated field and groundtruth. Top and bottom panels show plots of field estimation error versus receiver phase offset at two different SNRs. The fields were estimated using weighted least square regression (wLSR), penalized maximum likelihood (pML) and wLSR after estimation of absolute phase using LPE constraint (LPE+wLSR).

error as function of SNR in (a) and dependence of $\Delta TE$ in (b). At low SNR (SNR=4), it is seen that LPE+wLSR estimates the field with 66% reduction in error in comparison to wLSR and 56% reduction in error with respect to pML. As for the depentance on $\Delta TE$, it is observed that the LPE+wLSR estimated field do not exhibit significant increase in field estimation error with increase in $\Delta TE$.

## 3.2 Numerical Phantom experiments

For numerical phantom experiments, a volume was created in a $160 \times 160 \times 21$ matrix at $1 \times 1 \times 1$ mm$^3$ isotropic resolution. In this volume, small structures were embedded to mimic different brain nucleiwhich include caudate nucleus (CN), globus pallidus (GP), substantia nigra (SN), vein and paranasal sinus, with susceptibility values-8.8, -8.6, -8.4, -7.9 and 0 ppm, respectively. The susceptibility of white matter was chosen to be -9.0 ppm. Using this as the true susceptibility distribution $\chi(r)$, the field perturbation along the z-direction of the dipolar magnetic field was simulated using (28)

$$B_{dz}(k) = -\left(\frac{\chi(k)B_0}{3}\right)\left(3\frac{k_z^2}{k_x^2 + k_y^2 + k_z^2} - 1\right), \qquad [22]$$

where $\chi(k)$ is the Fourier transform of $\chi(j)$, $k_x, k_y$ and $k_z$ represent the k-space coordinates at $j$thposition. These field maps were then used to generate idealized multi-echo phase data that accumulates linearly over time with echo-time difference $\Delta TE = 6\,ms$. Complex data were then synthesized usingthe simulated phase and exponentially decaying magnitude with decay rate determined by prefixed value for $T_2^*$ at each location. Each magnitude and phase pair was then converted into real and imaginary images, to each of which a normally distributed zero mean noise with variance $\sigma^2$ was added to ensure comparable SNR in the simulated data. The noisy wrapped phase was then generated by computing the principal angle of the complex signal.

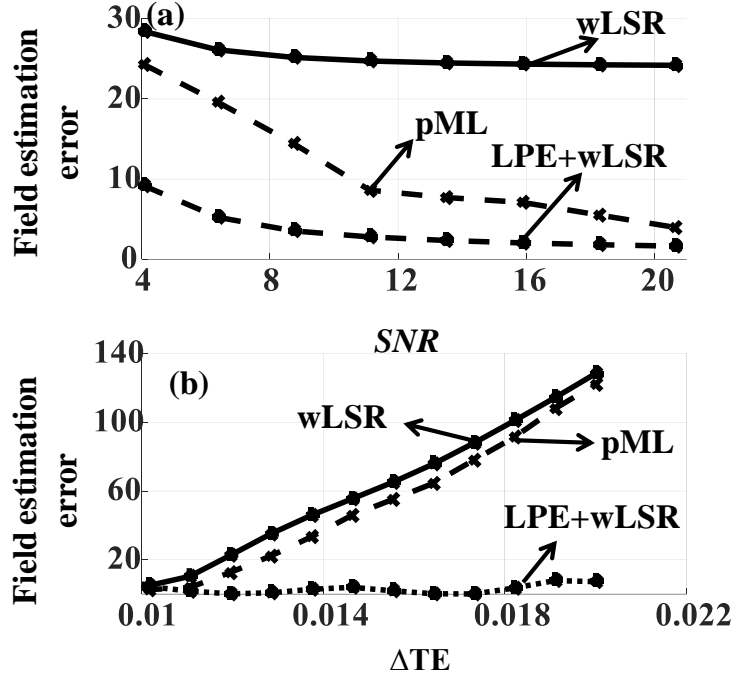

Figure 6: Field estimation error expressed as absolute difference between the estimated field and groundtruth. (a) plots of field estimation error versus SNR (b) plots of field estimation error versus ΔTE.

Fig. 7 below show the field maps estimated using wLSR, PML and LPE+wLSR. The left most panel (a) shows the groundtruth field map. The left-right panels show the field map obtained using (b) wLSR, (c) PML and (d) LPE+wLSR. The bottom row shows the field map estimation errors for the three methods. The relative-mean-square-error (RMSE) for each method is shown in the insets. It is observed that LPE+wLSR exhibit the minimum RMSE with respect to the groundtruth. For field map estimation, the relative reduction in RMSE for LPE+wLSR is found to be 98 % with respect to wLSR and 95% with respect to PML.

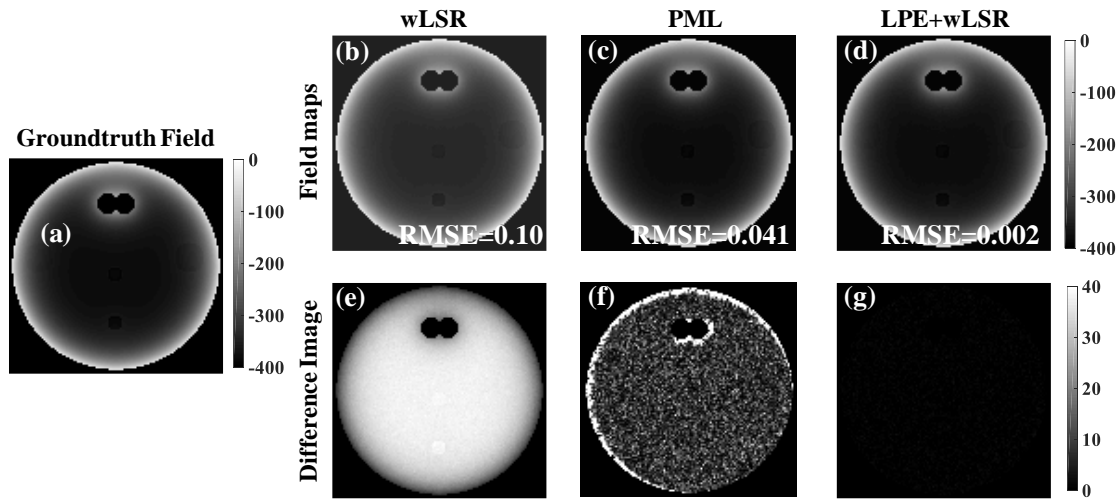

Figure 7: Field maps estimated using wLSR, PML and LPE+wLSR. The left most panel (a) shows the groundtruth field map. The left-right panels show the field map obtained from (b) wLSR, (c) PML and (d) LPE+wLSR. The bottom row show the field map estimation error expressed as the absolute difference between estimated field and groundtruth. The RMSE for each method is shown in insets.

## 3.2.1 Temporal phase unwrapping and Field estimation under sub-Nyquist condition

To simulate the sub-Nyquist condition along the temporal dimension, the field map was computed from Eq. [22] so as to satisfy the condition $\omega \Delta TE > \pi$ at $\Delta TE=12$ ms. Complex data were then simulated using exponentially decaying magnitudes and multi-echo phases as discussed in the previous section. Temporal phase unwrapping was then performed on the principal angles of the noisy complex image. Due to sub-Nyquist condition, Temporal unwrapping without prior rank approximation resulted in large errors in the unwrapped phase of the second echo as evidenced from the difference images shown in Fig. 8. Result of application of our unwrapping method in Section 2.5 after rank approximation, is shown in Fig. 8 (b3) and the corresponding difference image in Fig. 8 (c3).

The top row in Fig.8 shows the temporal variation in phase before and after temporal unwrapping at the locations highlighted using different colors. Panel (a1) shows the plot of wrapped 1D-phase and (a2) indicates the temporal unwrapped 1D-phase. Blue, red and yellow lines in (a2) indicates the ground truth, temporal unwrapped 1D-phase without rank approximation and the phase after rank approximation, unwrapped using our method in Section 2.5. The error maps expressed as the absolute difference between unwrapped phase of the second echo and the corresponding groundtruth are shown in panels (c2) and (c3). The last row shows the ground truth field (d1), the field estimated from temporal unwrapped phase without rank approximation (d2) and the field estimated using our method in Section 2.5 (d3). It is observed that this field is in close agreement with the ground truth field in (d1).

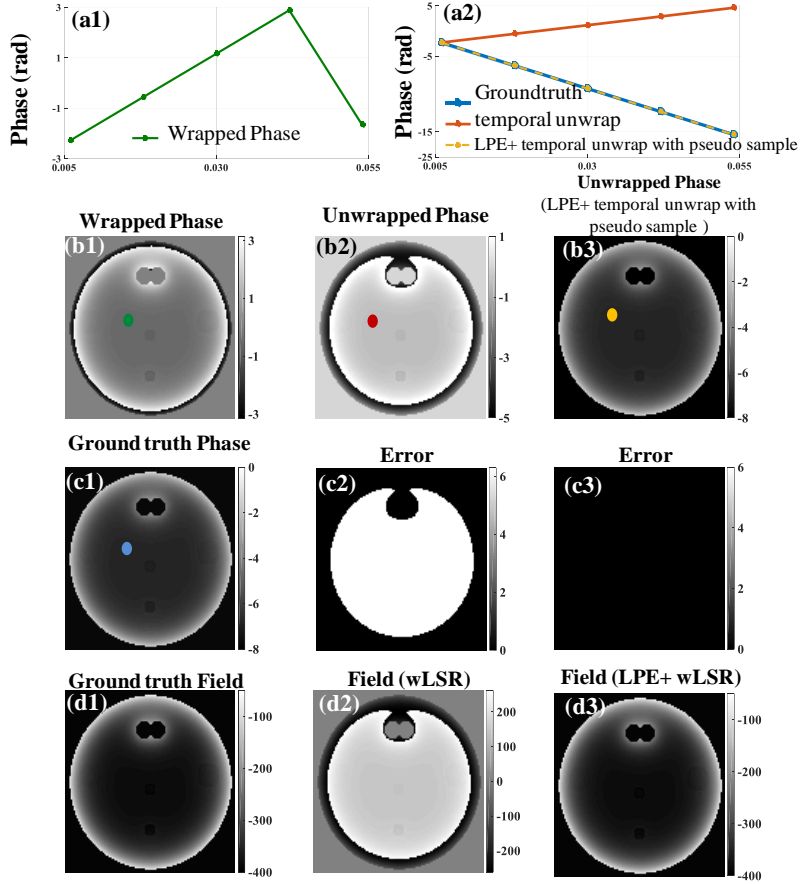

Figure.8: Temporal phase unwrapping under sub-Nyquist condition. Row-1: (a1) Wrapped phase along the temporal dimension at the location indicated in (b1), (a2) Plots of unwrapped 1D-phase. Row-2: (b1) Wrapped phase of second echo, (b2) Phase unwrapped without rank approximation and (b3) Phase unwrapped using proposed method in Section 2.5. Row-3: (c1) Ground truth phase of second echo, (c2)-(c3) Unwrapping error expressed as the absolute difference between the unwrapped phase and the groundtruth. Row-4: (d1) Groundtruth field, (d2) Field estimated from temporal unwrapped phase images of first and second echo without rank approximation and (d3) Field estimated from the phase images unwrapped using the proposed method in Section 2.5.

**Appendix**

The optimization problem of subproblem:-1 [16] is given by

$$\arg\min_{H_f} \frac{1}{2} \|D \circ H_f - G \circ H_h\|_F^2 + \frac{\rho}{2} \sum_{p,q=0}^{w-1} |D(p,q) \circ H_f(p,q) - d(p+q) \circ f(p+q) + u(p+q)|^2; \text{subject to rank}(H_x = 1). \quad [A1]$$

Letting $L_f(x, y) \triangleq u(p + q) - d(p + q) \circ f(p + q); 0 \leq p, q \leq w - 1$,

Eq.[A1] becomes

$$\arg\min_{H_f} \frac{1}{2} \|D \circ H_f - G \circ H_h\|_F^2 + \frac{\rho}{2} \sum_{p,q=0}^{w-1} |D(p,q) \circ H_f(p,q) + L_f(p,q)|^2;$$

$$\text{subject to } rank(H_f = 1).$$

[A2]

Since $\sum_{x,y=0}^{w-1} |D(p,q) \circ H_f(p,q) + L_f(p,q)|^2 \cong \|D \circ H_f + L_f\|_F^2$,

$$\arg\min_{H_f} \frac{1}{2} \|D \circ H_f - G \circ H_h\|_F^2 + \frac{\rho}{2} \|D \circ H_f + L_f\|_F^2;$$

$$\text{subject to } rank(H_f = 1).$$

[A3]

Using the definition $\|Z\|_F^2 = trace(ZZ^H)$,

$$\arg\min_{H_f} \frac{1}{2} trace\left[(D \circ H_f - G \circ H_h)(D \circ H_f - G \circ H_h)^H\right]$$

$$+ \frac{\rho}{2} trace\left[(D \circ H_f + L_f)(D \circ H_f + L_f)^H\right]$$

[A4]

Taking the derivative w.r.t $H_f$ and equating to 0,

$$[(D \circ H_f)D - (G \circ H_h)D] + \rho[(D \circ H_f)D + L_f D] = 0.$$  [A5]

then

$$(D \circ H_f) = [(G \circ H_h)D - \rho L_f D][D + \rho D]^{-1}.$$  [A6]

Since the RHS of [A6] and D are known quantities, $\widehat{H}_f^{k+1}$ can be estimated by dividing each element in the RHS of [A6] by the corresponding element in D.

Then $H_f^{k+1}$ can be obtained by performing truncated SVD and projecting this to the convex set of Hankel matrix.

$$H_f^{k+1} = \mathcal{D}(\widehat{H}_f^{k+1}), \qquad [A7]$$

where $\mathcal{D}$ represents the singular value thresholding (SVT) followed by convex projection. Here the projection is performed by averaging across the anti-diagonal elements (29) to restore the Hankel structure.

Similarly the optimization problem of subproblem:-2 [17] is given by

$$\arg\min_{f} \frac{\rho}{2} \sum_{p,q=0}^{w-1} \left| D(p,q) \circ H_f(p,q) - d(p+q) \circ f(p+q) + u(p+q) \right|^2 ; 0 \leq p,q \leq w-1. \qquad [A8]$$

On taking the derivative w.r.t $f$ and equating to 0,

$$D(p,q) \circ H_f(p,q) - d(p+q) \circ f(p+q) + u(p+q) = 0. \qquad [A9]$$

$$d(p+q) \circ f(p+q) = D(p,q) \circ H_f(p,q) + u(p+q); \qquad [A10]$$
$$0 \leq p,q \leq w-1.$$

Similar to the case in subproblem-1, $\hat{f}^{k+1}$ is obtained by dividing each element in RHS of [A10] with the corresponding element in $d$.

$$f^{k+1} = \mathcal{P}_C(\hat{f}^{k+1}), \qquad [A11]$$

where $\mathcal{P}_C$ defines the projection on to the convex set C of unit magnitude complex exponential signals. For an arbitrary complex exponential signal $Ae^{j\theta}$, this is implemented using $\arg\min_{\Phi} \|Ae^{j\theta} - e^{j\Phi}\|_2^2$, which after differentiation and equating to zero, gives $\Phi \triangleq \theta$.

# REFERENCES

1. Rauscher A, Sedlacik J, Barth M, Mentzel H-J, Reichenbach JR. Magnetic susceptibility-weighted MR phase imaging of the human brain. American Journal of Neuroradiology. 2005;26(4):736-42.

2. Lee J, Hirano Y, Fukunaga M, Silva AC, Duyn JH. On the contribution of deoxy-hemoglobin to MRI gray–white matter phase contrast at high field. Neuroimage. 2010;49(1):193-8.

3. Bilgic B, Pfefferbaum A, Rohlfing T, Sullivan EV, Adalsteinsson E. MRI estimates of brain iron concentration in normal aging using quantitative susceptibility mapping. Neuroimage. 2012;59(3):2625-35.

4. de Rochefort L, Liu T, Kressler B, Liu J, Spincemaille P, Lebon V, et al. Quantitative susceptibility map reconstruction from MR phase data using bayesian regularization: validation and application to brain imaging. Magnetic Resonance in Medicine: An Official Journal of the International Society for Magnetic Resonance in Medicine. 2010;63(1):194-206.

5. Liu T, Liu J, De Rochefort L, Spincemaille P, Khalidov I, Ledoux JR, et al. Morphology enabled dipole inversion (MEDI) from a single-angle acquisition: comparison with COSMOS in human brain imaging. Magnetic resonance in medicine. 2011;66(3):777-83.

6. Shmueli K, de Zwart JA, van Gelderen P, Li TQ, Dodd SJ, Duyn JH. Magnetic susceptibility mapping of brain tissue in vivo using MRI phase data. Magnetic Resonance in Medicine: An Official Journal of the International Society for Magnetic Resonance in Medicine. 2009;62(6):1510-22.

7. Schweser F, Deistung A, Lehr BW, Reichenbach JR. Quantitative imaging of intrinsic magnetic tissue properties using MRI signal phase: an approach to in vivo brain iron metabolism? Neuroimage. 2011;54(4):2789-807.

8. Deistung A, Schweser F, Wiestler B, Abello M, Roethke M, Sahm F, et al. Quantitative susceptibility mapping differentiates between blood depositions and calcifications in patients with glioblastoma. PloS one. 2013;8(3):e57924.

9. Barhoum S, Rodgers ZB, Langham M, Magland JF, Li C, Wehrli FW. Comparison of MRI methods for measuring whole-brain venous oxygen saturation. Magnetic resonance in medicine. 2015;73(6):2122-8.

10. Ward PG, Fan AP, Raniga P, Barnes DG, Dowe DL, Ng AC, et al. Improved quantification of cerebral vein oxygenation using partial volume correction. Frontiers in neuroscience. 2017;11:89.

11. Wehrli FW, Fan AP, Rodgers ZB, Englund EK, Langham MC. Susceptibility-based time-resolved whole-organ and regional tissue oximetry. NMR in biomedicine. 2017;30(4):e3495.

12. Jezzard P, Balaban RS. Correction for geometric distortion in echo planar images from B0 field variations. Magnetic resonance in medicine. 1995;34(1):65-73.

13. Sekihara K, Matsui S, Kohno H. NMR imaging for magnets with large nonuniformities. IEEE transactions on medical imaging. 1985;4(4):193-9.

14. Glover GH. Multipoint Dixon technique for water and fat proton and susceptibility imaging. Journal of Magnetic Resonance Imaging. 1991;1(5):521-30.

15. Windischberger C, Robinson S, Rauscher A, Barth M, Moser E. Robust field map generation using a triple-echo acquisition. Journal of Magnetic Resonance Imaging: An Official Journal of the International Society for Magnetic Resonance in Medicine. 2004;20(4):730-4.

16. Aksit P, Derbyshire JA, Prince JL, editors. Three-point method for fast and robust field mapping for EPI geometric distortion correction. 2007 4th IEEE International Symposium on Biomedical Imaging: From Nano to Macro; 2007: IEEE.

17. Kressler B, De Rochefort L, Liu T, Spincemaille P, Jiang Q, Wang Y. Nonlinear regularization for per voxel estimation of magnetic susceptibility distributions from MRI field maps. IEEE transactions on medical imaging. 2009;29(2):273-81.

18. Funai AK, Fessler JA, Yeo DT, Olafsson VT, Noll DC. Regularized field map estimation in MRI. IEEE transactions on medical imaging. 2008;27(10):1484-94.

19. Andersson F, Carlsson M, Tourneret J-Y, Wendt H. A new frequency estimation method for equally and unequally spaced data. IEEE Transactions on Signal Processing. 2014;62(21):5761-74.

20. Andersson F, Carlsson M, Maarten V. Sparse approximation of functions using sums of exponentials and AAK theory. Journal of Approximation Theory. 2011;163(2):213-48.

21. Goldstein RM, Zebker HA, Werner CL. Satellite radar interferometry: Two-dimensional phase unwrapping. Radio science. 1988;23(4):713-20.

22. Peng X, Ying L, Liu Y, Yuan J, Liu X, Liang D. Accelerated exponential parameterization of T2 relaxation with model-driven low rank and sparsity priors (MORASA). Magnetic resonance in medicine. 2016;76(6):1865-78.

23. Xu W, Qiao S. A fast symmetric SVD algorithm for square Hankel matrices. Linear Algebra and its Applications. 2008;428(2-3):550-63.

24. Jiang B, Ma S, Zhang S. Alternating direction method of multipliers for real and complex polynomial optimization models. Optimization. 2014;63(6):883-98.

25. Zhang Y. Restricted low-rank approximation via ADMM. arXiv preprint arXiv:151201748. 2015.

26. Boyd S, Parikh N, Chu E, Peleato B, Eckstein J. Distributed optimization and statistical learning via the alternating direction method of multipliers. Foundations and Trends® in Machine learning. 2011;3(1):1-122.

27. Gabay D, Mercier B. A dual algorithm for the solution of nonlinear variational problems via finite element approximation. Computers & mathematics with applications. 1976;2(1):17-40.

28. Marques J, Bowtell R. Application of a Fourier-based method for rapid calculation of field inhomogeneity due to spatial variation of magnetic susceptibility. Concepts in Magnetic Resonance Part B: Magnetic Resonance Engineering: An Educational Journal. 2005;25(1):65-78.

29. Cai J-F, Candès EJ, Shen Z. A singular value thresholding algorithm for matrix completion. SIAM Journal on optimization. 2010;20(4):1956-82.